\documentclass[preprint,aip,jcp,floatfix]{revtex4-1}

\usepackage{hyperref} 
\usepackage{graphicx} 
\newcommand{\e}{$|e^-|$}

\begin{document} 

\title{Partial Ionic Bonding in Homogeneous Sodium Clusters}

\author {Vaibhav Kaware}
\affiliation{Department of Physics, University of Pune, India - 411007}
\affiliation{Physical and Materials Chemistry Division, CSIR-National Chemical
Laboratory, Pune, India - 411008}
\author {Kavita Joshi}
\email{k.joshi@ncl.res.in,kavita.p.joshi@gmail.com} 
\affiliation{Physical and Materials Chemistry Division, CSIR-National Chemical
Laboratory, Pune, India - 411008}

\begin{abstract}
In this work, we report an interesting observation of partial ionic 
bonding due to charge transfer in \textit{homogeneous} sodium clusters. The charge
transfer causes the electronic charge to accumulate on the surface, and the
resulting charges on atoms range between +0.4 to -1.0 \e.
We also demonstrate that this disparity among ef\mbox{}fective charges on atoms
is geometry dependent, such that atoms experiencing similar surrounding, have
equal ef\mbox{}fective charge.
It is speculated that 
this phenomenon will occur among other homogeneous clusters as well, and its
extent will be def\mbox{}ined by the valence electron delocalization. 
\end{abstract} 

\maketitle
\section{Introduction}
Atomic clusters are interesting, useful, and
intriguing.\cite{catjenapnas,catpnas,aunaturecatreview,
aunaturebioreview,auforco,rhprl,jarroldjacs,catkhannascienceal,
snhightm,toohottomelt,gajacs,gongga,issendorfreview,gamagicgeometric,gahightm} 
Their f\mbox{}inite size 
has brought out many counterintuitive observations about their various properties. 
For instance, bulk gold, which is a noble metal, acts as a catalyst in its cluster
form. It is used as a chemical catalyst, as well as in biomedicine.
\cite{aunaturecatreview, aunaturebioreview,auforco} 
Bulk rhodium, which is a non-magnetic solid, turns magnetic in its cluster form.
\cite{rhprl}
Aluminum, which is usually used in its compound form as catalyst, shows
catalytic activity when used as clusters.\cite{jarroldjacs,catkhannascienceal}
Some clusters also exhibit higher than bulk melting
temperature, which contradicts our understanding that melting temperature 
lowers with decreasing size.\cite{snhightm,toohottomelt,gajacs}
All these odd properties of clusters are dependent upon the structure as well as
their bonding.
Bonding in clusters is quite dif\mbox{}ferent than their corresponding bulk
material, and changes even across sizes of clusters of the same element. 
Gallium in its bulk form has metallic as well as covalent type bonding.
\cite{gongga}
Bonding in its clusters, however has been an unresolved key issue.
\cite{gahightm,throwing,aguado2012nanoscale}
Lead clusters are agreed to change their bonding from metallic to non-metallic,
but the size at which this occurs, is still under debate.
\cite{pbmetaltononmetal,pbcovalentmetallic,largepbclusters} 
Tin shows a semiconductor to metal transition with changing size of its
clusters.\cite{snsemitometalwang}
Thus, bonding among clusters has been a fairly debatable issue for 
clusters of many elements, and is a key factor in understanding and
explaining their behavior.
Amidst all these controversies, bonding in sodium clusters has come clean till
now. Sodium clusters have been shown to possess metallic bonding.
\cite{howmetallicna2,howmetallicna,howmetallicnanature,issendorfreview} 
In this work, we show that bonding in sodium clusters is more than just metallic,
using atomic charge measurements.
Charge transfer in heterogeneous systems
is a known fact, where the amount of charge that gets transferred, depends upon the 
electron af\mbox{}f\mbox{}inities of corresponding elements. 
In this work, we present an evidence of substantial charge transfer observed 
in \textit{homogeneous} clusters of sodium. Such charge transfer 
is known to occur, in small quantities, in a few homogeneous 
clusters.\cite{b80prl,catkhannascienceal,sailajaau6}
However, to the best of our knowledge, its extent 
and its influence on bonding have not been studied to date.

In what follows, we discuss the charge distribution 
of homogeneous sodium clusters with sizes up to few hundred atoms. 
We present Bader charge analysis of sodium clusters with
sizes ranging from  few 10 to few 100 atoms.
Specif\mbox{}ically, we demonstrate that in neutral sodium clusters,
most of the atoms either gain or loose charge, and that the excess charge on 
each atom varies from +0.4 to -1.0 \e. 
Thus, in a cluster, some atoms are positively charged, while
some others possess an ef\mbox{}fective -ve charge. This gives rise to a somewhat
counterintuitive phenomenon 
of partial ionic bonding among homogeneous clusters of sodium atoms.
As a result of the charge transfer, 
electronic charge is observed to accumulate on the surface of the cluster and 
towards center. 
Geometry plays a crucial role and atoms with
identical environment have identical ef\mbox{}fective charge. 
In case of highly symmetric clusters, like 55 and 147, we observe that atoms 
belonging to the same shell have identical charge polarity. 
Charge dif\mbox{}ference is more 
for geometries that are symmetric and ordered.
By `ordered', we mean geometries with more number of atoms experiencing similar
environment, in terms of neighboring atoms.
Although it is possible to infer and correlate our
observations with polarizability and electric dipole moment, it should be noted
that the focus of this work is upon bringing out the interesting observation about
structure induced inhomogeneity of charge distribution in homogeneous sodium clusters. 
The article is organized as follows.
Section \ref{sec:compute} outlines the  computational details, 
and section \ref{sec:results}, 
the results and discussions. 
Conclusions are drawn in section \ref{sec:concl}.

\section{\label{sec:compute}Computational Details}
 Ab initio molecular dynamics is carried out within the Kohn-Sham formulation
 of density functional theory (DFT). Projector Augmented Wave pseudopotential
 \cite{paw1,paw2} is used, with Perdew Burke Ehrzenhof 
 (PBE) \cite{pbe1} approximation for the exchange-correlation and generalized gradient
 \cite{pbe2} approximation, as implemented in
 plane wave code, Vienna Ab initio Simulation Package (VASP).\cite{vasp1,vasp2,vasp3} 
 Energy cutof\mbox{}f for plane-waves is kept at 102 eV for f\mbox{}inite temperature
 molecular dynamics. 
 Relaxation runs are carried with high precision
 setting in the VASP package, raising the ef\mbox{}fective energy cutof\mbox{}f to 127 eV. 
 Cubic simulation cell, with image in each cell separated by at least 10 \AA, is used.
 Energy convergence criterion of 10$^{-5}$ eV, and a force cutof\mbox{}f of 0.001 eV/\AA~
 are used for relaxation.  
The geometries used in this work are obtained
from the authors of previous works,\cite{nastructures} and 309 and 561 icosahedra are 
taken from the Cambridge Cluster Database.\cite{ccdb}
All the geometries used in this work are relaxed within def\mbox{}ined
computational accuracy, at the level of theory used here.

We have carried out Bader charge analysis for sodium clusters of numerous sizes ranging 
from 10-80, intermittent sizes up to 147, and larger icosahedral clusters with 309
and 561 atoms.  
Bader charge analysis was chosen on account of its reliability in comparison 
with other methods like Mulliken, Hirshfeld, that are 
used to calculate atomic
charges.\cite{baderreliability,badercompare,baderhfeld,baderurl}
Bader analysis makes use of the fact, that
in molecular systems, the charge density between atoms reaches a minimum.
The point of minimum charge density is used to separate atoms from each other.
This analysis def\mbox{}ines atoms using a 2-D surface, on which the charge density 
is a minimum, perpendicular to the surface.
The charge enclosed within the Bader volume is a good approximation to the
total electronic charge of an atom.\cite{bader1,bader2,bader3}
In order to be able to reproduce total charge on system correctly, 
f\mbox{}ine meshes are used for Bader charge calculations. 
The same f\mbox{}ine FFT meshes are used for computing wavefunctions and charges during
relaxation.
We make use of 2-D mapping of atoms, described below, to analyze the occurring 
redistribution of charges.
\subsection{\label{sec:twodmapping}2-D mapping of atoms}
 In order to quantify the ef\mbox{}fective charge on each atom as a function of
 its position, each atom is mapped onto the 2-D plane and is colored according 
 to the charge it possesses. 
 For this mapping, we
 plot the perpendicular distance of each atom from the three axes, 
 $x$, $y$, and $z$, against the $x$, $y$ and $z$ coordinate of that
 atom. 
 For better graphical representation, the distance of each
 atom from $x$, $y$ and $z$ axes is given the
 arithmetical sign of $y$, $z$ and $x$ co-ordinate of that atom. 
 Such a representation of atoms from 3-D to 2-D is most useful, when dealing
 with spherical geometries. It helps bringing out the shells in spherical
 geometries, whenever present.
 Each point on this 2-D graph represents an atom, and is colored according 
 to the charge it possesses. 
\subsection{\label{sec:chargeconv}Charge measuring convention}
\begin{figure}
  \includegraphics[scale=0.6]{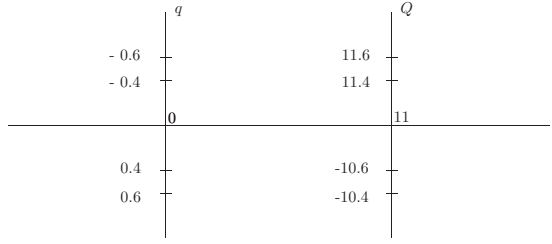}
\caption{Diagram showing the convention of measuring ef\mbox{}fective charge 
used in this work, in comparison with the total charge on sodium atom. 
$q$ ~: Ef\mbox{}fective charge on sodium atom as measured in this work. 
$Q$ : Actual charge on sodium atom.}
\label{fig:chargeconv}
\end{figure}

 Charge on each atom is measured in excess of 1 valence electronic charge of a
 neutral Na atom. Schematics of the convention are shown in Fig. \ref{fig:chargeconv}. 
 For instance, any atom with charge greater than 1.0 \e, is given a -ve sign (since it has
 accumulated -ve charge), and a magnitude in units of `excess of 1.0 \e'. 
 The charge polarity is also def\mbox{}ined
 based on the same scale. For example, any atom gaining 0.4 \e~excess electronic 
 charge is said to have a -0.4 \e~charge, while an atom loosing 0.4 \e~electronic
 charge, is said to possess +0.6 \e~charge.
\subsection{\label{sec:shapedefine}Shape parameter}
Shape parameter $\varepsilon_{\mathrm{def}}$ is used to quantify the distribution of
atoms along the three spatial directions.\cite{epsdef} It is a rough measure to quantify the
overall shape of the cluster. It is def\mbox{}ined as:
\[
 \varepsilon_{\mathrm{def}}=\frac{2Q_x}{Q_y+Q_z},
\] where, $Q_x\geq Q_y\geq Q_z$ are the eigenvalues of the quadrupole tensor
def\mbox{}ined as $Q_{ij}=\sum_I R_{Ii} R_{Ij}$. $R_{Ii}$ stands for the $i^{~th}$
coordinate of ion $I$, measured from the origin. 
The structure is prolate, when
$Q_x \gg Q_y\approx Q_z$. 
For a spherical geometry, $Q_x = Q_y = Q_z$, and $\varepsilon_{\mathrm{def}}=1$.
\section{\label{sec:results}Results and Discussion}
\begin {figure}
  \centering
  \includegraphics[scale=0.58]{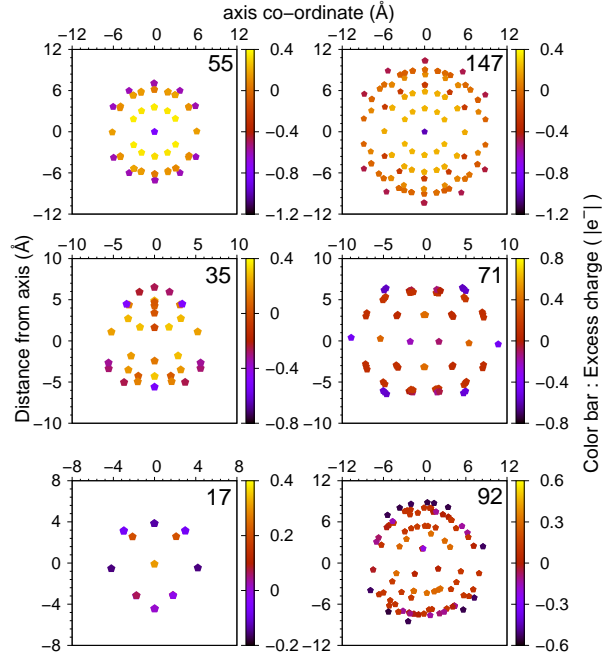}
  \caption{Charge distribution for various Na clusters. Electronic charge
  accumulates on the surface of all these clusters. Atoms at center also
  accumulate -ve charge, except that of 17. Spatial symmetry of clusters gets
  reflected in charge distribution.}
  \label{fig:chargedistriball}
\end {figure}
We begin our results and discussion by presenting 
2-D mappings of atoms, colored by magnitude of ef\mbox{}fective charge on each atom, 
for various sizes, in Fig. \ref{fig:chargedistriball}. 
Cluster geometries for these sizes, are shown in Ref.\cite{SI}.
The sizes of clusters are mentioned at the top right of each frame.
The axes in the graph represent the position of each atom with respect to center
of mass (COM), in 2-D mapping.
The f\mbox{}irst two f\mbox{}igures are for sizes 55 and 147. Both these structures are 
Mackay icosahedra.
In these structures, -ve charge clearly gathers on the surface, and on the central
atom.
These two clusters comprise of atoms arranged in concentric shells, and as seen
from the f\mbox{}igure, charge on atoms that belong to the same shell has same polarity
and hardly varies in magnitude. This is attributed to the fact
that atoms in the same shell confront identical environment.
Same is further demonstrated in geometries 35 and 71. These two geometries are
not concentric shells, but have local order. 
35 has three centers of local order, and that shows up here as three dif\mbox{}ferent
sets of atoms possessing identical charge distribution.  Similarly,
71 is a structure formed by two  intermeshing size 55 icosahedra with
two centers of local order. Owing to this two centered local
order, we expect a two centered charge distribution. 
The graph for size 71 does show two innermost atoms possessing high electronic 
charge (dark shade). This is in line with the fact that electronic charge prefers 
to accumulate towards the center of cluster. 
Size 17 is the smallest size with structure that encloses a sodium atom. 
It exhibits a dif\mbox{}ferent behavior than other sizes, which is due to its
small size. Particularly, its central atom has ef\mbox{}fective +ve charge, in contrast
with other sodium clusters. Its surface has an overall -ve charge, 
with 10 out of 16 atoms charged -ve, four charged +ve, and the rest neutral.
The last frame in the f\mbox{}igure is for size 92. The ground state (GS) for this size
is reported to
possess a non-icosahedral local order,\cite{localorderna} 
which is reflected in
the charge distribution shown here. The four atoms at the innermost shell have
almost identical charge. Also, with a couple of exceptions, all other
atoms in their respective shells have little or no variation in magnitude of 
charge that they posses (maximum variation in magnitude of charges in same 
shell is $\sim$0.03 \e).
This can be attributed to the local order that 92 possesses. 
Thus, we demonstrate that various dif\mbox{}ferent geometries of sodium clusters
show the common trait that free standing sodium clusters possess 
a -vely charged surface, with `ef\mbox{}fective charge on an atom' being a function of
its neighborhood.

\begin {figure} 
  \includegraphics[scale=1.00]{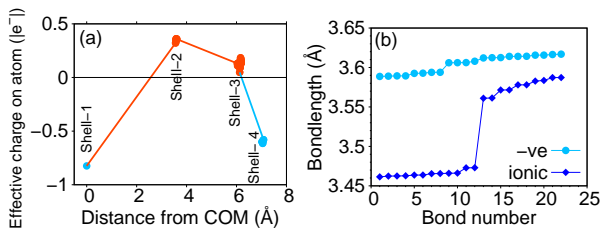}
  \caption{Analyses for Na$_{55}$: (a) All atoms in a given shell have same
  charge polarity.
  (b) First 22 shortest bonds in Na$_{55}$ are between atoms with opposite
  polarity (dark blue). Shortest bondlengths between atoms with -ve polarity 
  are shown in light blue.}
  \label{fig:na55}
\end{figure}
\begin{figure}
  \includegraphics[scale=0.60]{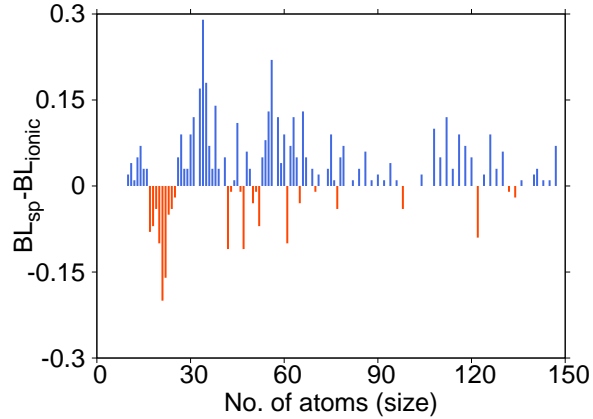}
  \caption{Plot of dif\mbox{}ference between the shortest ionic bondlengths, and the 
  shortest bondlengths
  of atoms with same polarity, plotted against the size of the cluster. 
  The shortest bonds are between atoms with opposite polarity for most of 
  the sizes.}
  \label{fig:ionicbldiff}
\end{figure}

Charge transfer among atoms of Na$_{55}$ is further investigated in detail.
A 55 atom icosahedron consists of 4 concentric
shells with 1, 12, 30 and 12 atoms. 
Fig. \ref{fig:na55}(a)
shows the ef\mbox{}fective charge on each atom, plotted
against its distance from center of mass (DCOM).  
Atoms that gain electronic charge, are colored blue,
while those that loose electronic charge, are colored orange in the plot.
We see from the f\mbox{}igure that
atom at the COM gathers an excess of 0.8 \e~electronic charge, 
while all
atoms in the second shell (DCOM=3.61 \AA) loose about 0.4 \e~electronic charge each.
Similarly, atoms in the third shell loose electronic charge and the outermost
shell atoms gain it. The collective magnitude of this charge transfer is quite 
large for the  entire shell of atoms.
There is a change of charge polarity accompanied by a nearly 11 \e~
cumulative charge dif\mbox{}ference, between
the third and the fourth shell, as well as between second and 
the fourth shell. 
Hence, we speculate that this transfer of
charge among the shells leads to an ionic contribution to the stability of
bonds formed between atoms of these shells.
Further analysis of bondlengths conf\mbox{}irms this speculation. It
shows that the shortest bonds are formed between atoms of 
the 4$^{th}$ and the 2$^{nd}$ shell. 
This indicates that the shortest bonds in this cluster 
posses an ionic character to it, owing to disparity of charges on respective atoms.
Plotting the bondlengths of atoms of same polarity, and
opposite polarity, shows that the set of f\mbox{}irst 22 shortest bonds 
are those formed between atoms with opposite polarity, as shown in Fig.
\ref{fig:na55}(b).
This comparison was also made for all 107 dif\mbox{}ferent sizes between 10-147. 
It revealed that 84 out of these 107 sizes had their shortest bond formed between
atoms with opposite polarity (ionic).
Fig. \ref{fig:ionicbldiff} shows the plot of dif\mbox{}ference between the shortest bond
between atoms with same polarities (BL$_{\text{sp}}$) 
and the shortest bond between atoms with opposite
polarities (BL$_{\text{ionic}}$).
The dif\mbox{}ference is positive whenever the shortest bond is ionic, and negative
otherwise. As seen from the f\mbox{}igure, most of the sizes show this dif\mbox{}ference to be
positive, which implies their shortest bonds are between atoms with opposite
polarity.
This clearly demonstrates the overall influence of ionic character on
bonding in sodium clusters, 
and that the ionic contribution to bonding in sodium clusters 
cannot be overlooked within this size range.

\begin {figure}
  \includegraphics[scale=1.00]{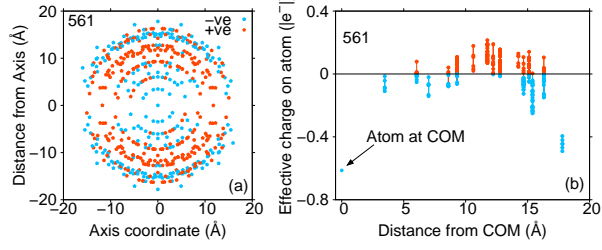}
  \caption{(a) Distribution of charge polarities of atoms in each shell
  of Na$_{561}$.
  (b) Charge on each atom as a function of its distance from center 
  of mass for Na$_{561}$. For larger sizes, atoms in same
  shell do not posses identical charge polarity. Also, the variation in charge
  magnitudes is greater in each shell, in comparison with smaller icosahedra.
  }
  \label{fig:chargedistribtoggle}
\end {figure}

Next, we compare observations noted for Na$_{55}$, with higher sized icosahedra.
As shown in Fig. \ref{fig:chargedistribtoggle}, 
with increasing number of atoms in each shell, polarity of
ef\mbox{}fective charge on each atom in a given shell does not remain same. 
Fig. \ref{fig:chargedistribtoggle}(a) consists of the 2-D mapping of 
atoms for
size 561. The points on this graph are colored by 
polarity of ef\mbox{}fective charge on each atom, and not their magnitude. 
Atoms that gain electronic charge, are colored
blue, while those that loose electronic charge are colored orange.
The f\mbox{}igure makes it clear that there is an electronic charge 
accumulated towards center and on the surface. 
Also, the uniform charge polarity and magnitude among all atoms in same
shell, previously seen in Na$_{55}$, is absent here. 
This is shown in Fig. \ref{fig:chargedistribtoggle}(b), in which  magnitude of 
charge transfer on each atom is plotted as a function of the respective DCOM.
We also note that variation in ef\mbox{}fective charge on each atom in a given shell, 
is more compared to smaller icosahedra. 
Similar analyses for size 147 and 309 atom icosahedra, are presented in Ref.
\cite{SI}. 
Although 147 icosahedron follows same
trends as that of 55, 309 icosahedron marks the beginning of trends observed in 
561.

\begin {figure}
      \includegraphics[scale=0.62]{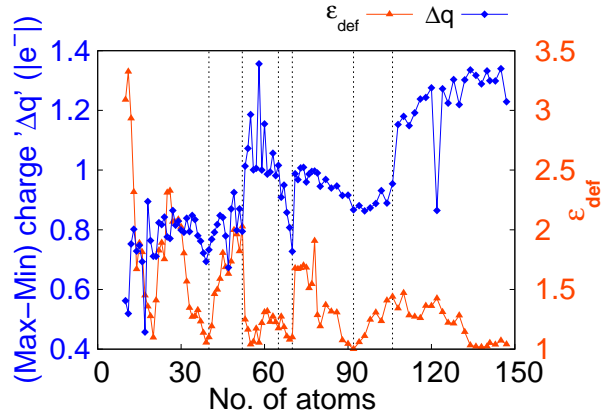}
   \caption{Dif\mbox{}ference between maximum and minimum charge on an atom in 
   cluster ($\Delta$q), as a function of size of the cluster. 
   Right $y$-axis shows variation in shape parameter
   ($\varepsilon_{\mathrm{def}}$). 
   $\Delta$q follows the changes in geometrical motifs  closely.
   }
   \label{fig:chargediffsize}
\end {figure}
While the ef\mbox{}fective charge on atoms should show no variation 
for sodium bulk, it is interesting to study the same for clusters, as a
function of their size and shape.
Hence, variation of the `dif\mbox{}ference between maximum and minimum 
ef\mbox{}fective charge on an atom in a cluster'
($\Delta$q) is studied as a function of size of the cluster.
Fig.  \ref{fig:chargediffsize} shows a plot $\Delta$q as a function of size 
of the cluster. Shape deformation parameter $\varepsilon_{\mathrm{def}}$ 
is plotted on the right $y$-axis
of graph.
The f\mbox{}igure shows that $\Delta$q is maximum whenever the geometry assumes
icosahedral motif.
Sodium clusters show globally disordered structures with local icosahedral order
starting from size 19 intermittently till 52. A sudden change in motif occurs
from 52 to 53.
Icosahedral order is then followed for sizes 53 to 64. 
$\Delta$q also shows the sudden jump from 52 to 53, and has high value till 
size 64.
The ground state structures are disordered between 64 and 70, and 
$\Delta$q also shows a dip in
its value in this size range. Change of motif from one centered to
two centered icosahedra is indicated by sudden rise in $\varepsilon_{def}$ at 70.
$\Delta$q mimics the same, and goes hand in hand with this change of motif.
Clusters become more disordered with less icosahedron like structure till 92 and
up to 106, gradually. $\Delta$q also lowers in value gradually between 71 through
92 and up to 106. While value of $\varepsilon_{\mathrm{def}}$ is very close to 1 
around size 90, the order is not icosahedral.
This correlates well with observed variation in $\Delta$q. 
After size 106, the geometrical order
regains the icosahedral motif, which builds up and completes the 
next icosahedron at 147. Similar growth trend is followed by $\Delta$q as well.
Sudden drop in $\Delta$q at size 122 is due to abrupt displacement of its
atom nearest to COM at 2.5\AA, in comparison with its neighboring sizes, who have
nearest atom from their COM at 1.5\AA. 
Thus, we see that $\Delta$q follows the growth pattern of
sodium clusters closely, and maximizes whenever the geometries have icosahedral
motif. For larger sizes 309 and 561, $\Delta$q has lower values 
(1.24 and 0.82 \e), which points to the fact that as the system goes on becoming
larger, this quantity dies out. It will reach the value of zero for inf\mbox{}inite
solid bulk sodium. This is in agreement with our understanding that in solid
bulk, all sodium atoms should possess identical charge, on account of the identical
surrounding of each of them.

\begin {figure}
      \includegraphics[scale=0.62]{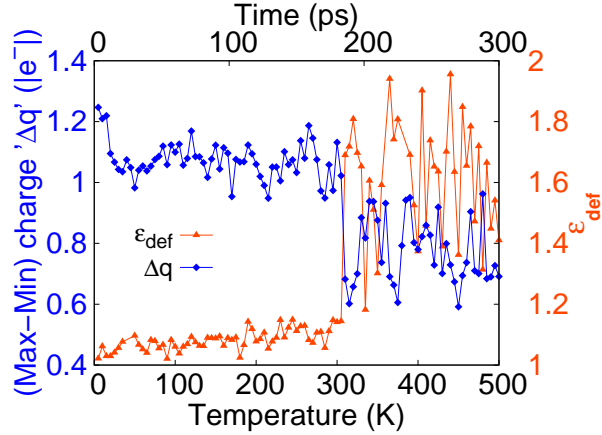}
   \caption{$\Delta$q for
    100 dif\mbox{}ferent geometries sampling a heating run of Na$_{55}$ between 0-500
    K. Right $y$-axis is the corresponding variation in shape parameter 
    ($\varepsilon_{\mathrm{def}}$). Out of phase variation in $\Delta$q and
    $\varepsilon_{\mathrm{def}}$ signif\mbox{}ies $\Delta$q's preference for spherical
    geometries.
    }
  \label{fig:chargediffna55heat}
\end {figure}
Next, we investigate the ef\mbox{}fect of temperature on 
charge transfer as observed in Na$_{55}$. 
Towards this end, we have performed slow-heating of Na$_{55}$. 
MD run was started with initial geometry as the icosahedron.
100 dif\mbox{}ferent geometries are picked out of this heating run carried
out between 0 to 500 K, in 300 ps. 
The sampling is done in an unbiased way, by picking out a geometry every 3 ps. 
Bader analysis is performed for these unrelaxed geometries in order to
investigate charge redistribution at f\mbox{}inite temperature.
Fig. \ref{fig:chargediffna55heat} is a graph of $\Delta$q for these geometries.
In this f\mbox{}igure, temperature is plotted on $x$ axis, 
and the simulation time on upper $x$-axis. 
Right $y$-axis shows the corresponding variation in shape parameter
($\varepsilon_{def}$).
Value of $\Delta$q is the highest in the graph initially, when the geometry is a
perfect icosahedron. It drops initially, as the cluster changes shape due to
increased temperature, as atoms begin to vibrate about the ideal icosahedron
positions. $\Delta$q values oscillate about 1.05 \e~during the phase when
cluster is distorted, before it actually melts. $\Delta$q suf\mbox{}fers another drop
upon melting of cluster, and even after melting, it attains higher values
whenever the cluster becomes spherical. 
In this f\mbox{}igure, high values of $\Delta$q occur at points when 
$\varepsilon_{def}$ shows a low
value (close to 1). This implies that $\Delta$q is large for spherical
geometries, and has lower values otherwise.

Based on all the analyzes mentioned above, we have brought out a 
simple, yet counterintuitive and interesting
picture of bonding in homogeneous sodium clusters.
Indeed, it is an artifact of the f\mbox{}inite 
size of the system.
In sodium bulk, 
all the atoms witness identical environment, 
and there is no reason for its the delocalized
electrons to favor certain atomic sites in preference over the others.
All atomic sites in the inf\mbox{}inite solid are equivalent in terms of their 
neighboring environment.
In contrast to this, in a f\mbox{}inite sized system, most atoms hold unique
positions in terms of their neighboring environment, along with 
existence of certain equivalent
positions, that are decided by the symmetry of the cluster.
Atoms at the distinct positions have distinct ef\mbox{}fective charge, while
atoms experiencing identical environment possess identical ef\mbox{}fective charge.
An interesting manifestation of this is seen in icosahedral structures for
various sizes from 55 to 561.
Thus,
all atoms do not experience equivalent environment in a cluster, which results
into partial ionic bonding in these clusters. 
Hence, we see that this
counterintuitive phenomenon of partial ionic bonding among a f\mbox{}inite collection
of homogeneous atoms, is an artifact of the f\mbox{}inite size of the system, and
should be expected to occur among all f\mbox{}inite sized system to a varied degree. 

\section{\label{sec:concl}Conclusions}
We have demonstrated that charge transfer occurs in homogeneous
sodium clusters, which gives rise to ionic character in bonding. 
Charge tends to accumulate onto the surface of these clusters, and
towards their center. 
Magnitude of ef\mbox{}fective charges depends on the size as well as symmetry of 
the cluster. The ef\mbox{}fect is most prominent for clusters with few 10s to few 100s
of atoms.
The charge dif\mbox{}ference between atoms is more for more symmetric
clusters.
Since the charge transfer observed here is geometry and
symmetry dependent, one may expect this phenomenon to occur in
homogeneous clusters of other elements as well, in varying magnitudes. 

\section{Acknowledgements}
Authors would like to acknowledge the Council of Scientif\mbox{}ic and Industrial
Research for f\mbox{}inancial support (Project No. CSC-0129 and CSC-0128). 
VK is grateful to the Department of Science and Technology (Project No. GOI-555A) for
partial f\mbox{}inancial support.
The authors thank S. M. Ghazi for providing sodium cluster geometries.

\bibliography{bibliography}
\bibliographystyle{unsrt} 
\end{document}